\pdfoutput=1

\documentclass[11pt]{article}

\usepackage[preprint]{acl}

\usepackage{times}
\usepackage{latexsym}
\usepackage{booktabs}
\usepackage{amsmath}
\usepackage{forest}
\usepackage{tikz}
\usepackage{float}
\usepackage{amsfonts}
\usepackage{multirow}
\usepackage{cleveref}

\usepackage[T1]{fontenc}

\usepackage[utf8]{inputenc}

\usepackage{microtype}

\usepackage{inconsolata}

\usepackage{graphicx}

\usepackage{forest}
\usepackage{graphicx}
\usepackage{amsmath}
\usepackage{paralist}
\usepackage{url}

\title{Augment or Not? A Comparative Study of Pure and Augmented Large Language Model Recommenders}

\author{Wei-Hsiang Huang$^*$\quad Chen-Wei Ke$^*$\quad Wei-Ning Chiu$^*$\quad Yu-Xuan Su \\ \textbf{Chun-Chun Yang \quad Chieh-Yuan Cheng\quad Yun-Nung Chen\quad Pu-Jen Cheng}\\
  National Taiwan University, Taipei, Taiwan \\
  \texttt{\{r13944004,r13944029,r12922219,r13946008,r13944037,r13944023\}@ntu.edu.tw} \\  \texttt{y.v.chen@ieee.org} \quad \texttt{pjcheng@csie.ntu.edu.tw} }

\newcommand{\uset}{\ensuremath{\mathcal{U}} }
\newcommand{\iset}{\ensuremath{\mathcal{I}} }
\newcommand{\mset}{\ensuremath{\mathcal{M}} }
\newcommand{\rset}{\ensuremath{\mathcal{R}} }

\crefformat{section}{#2Section~#1#3}
\crefformat{subsection}{#2Section~#1#3}
\crefformat{subsubsection}{#2Section~#1#3}
\crefformat{figure}{#2Figure~#1#3}
\crefformat{table}{#2Table~#1#3}
\crefformat{appendix}{#2Appendix~#1#3}

\begin{document}

\maketitle
\begin{abstract}
Large language models (LLMs) have introduced new paradigms for recommender systems by enabling richer semantic understanding and incorporating implicit world knowledge.
In this study, we propose a systematic taxonomy that classifies existing approaches into two categories: (1) Pure LLM Recommenders, which rely solely on LLMs, and (2) Augmented LLM Recommenders, which integrate additional non-LLM techniques to enhance performance.
This taxonomy provides a novel lens through which to examine the evolving landscape of LLM-based recommendation.
To support fair comparison, we introduce a unified evaluation platform that benchmarks representative models under consistent experimental settings, highlighting key design choices that impact effectiveness.
We conclude by discussing open challenges and outlining promising directions for future research. This work offers both a comprehensive overview and practical guidance for advancing next-generation LLM-powered recommender systems.\footnote{\url{https://github.com/MiuLab/LMRec-Survey}}
\begingroup\def\thefootnote{\rm *}\footnotetext{Equal contribution.}\endgroup
\end{abstract}

\section{Introduction}

The emerging capabilities of LLMs such as BERT~\cite{LLM:BERT}, T5~\cite{LLM:t5}, Llama~\cite{LLM:llama2} and GPT~\cite{LLM:GPT3} have sparked growing interest in their application to recommendation systems. Unlike traditional approaches that primarily rely on structured user-item interaction patterns,
leveraging semantic understanding and reasoning capabilities of LLMs offers a new paradigm for recommendation systems.
By incorporating rich textual metadata and the implicit knowledge encoded in LLMs, LLM Recommenders provide an opportunity to enhance performance and mitigate common recommendation challenges, such as cold-start issue and cross-domain generalizability problem, discussed in~\cref{sec:future}.

Besides leveraging LLMs solely as recommenders, recently, a new wave of research~\cite{Paper:31:Llara,Paper:25:TIGER,Paper:21:RPP} is now exploring to integrate non-LLM techniques, such as conventional collaborative models~\cite{Traditional:GRU4Rec,Traditional:SASRec} or clustering methods~\cite{Others:RQ-VAE}, to help the final decision making of LLM Recommenders.

\begin{figure}[t]
  \centering
  \includegraphics[width=0.5\textwidth]{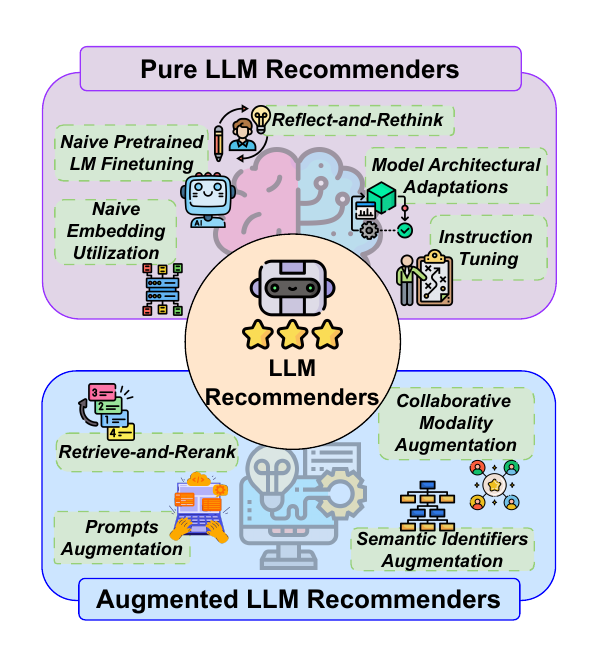}
  \vspace{-2mm}
  \caption{\textbf{An illustration of the taxonomy.} LLM Recommenders can be categorized into Pure (up) and Augmented (down) LLM Recommenders, depending on whether they utilize non-LLM techniques to help the final decision making of LLMs.}
  \label{fig:illustration_methods}
\end{figure}

With the growing interest and parallel development in both pure LLM-based approaches and those augmented with non-LLM techniques, it is crucial to systematically understand the different aspects of both scenarios.
Accordingly, unlike previous surveys~\cite{Related_Survey:1, wu2024survey, Related_Survey:2}, we present a fresh categorization of methods based on whether they incorporate non-LLM techniques to augment LLMs in making final recommendation decisions. This perspective aligns with the emerging trend of integrating LLMs into recommendation systems. An illustration is given in~\cref{fig:illustration_methods}.
Furthermore, we compare these approaches through dedicated experiments conducted on a unified evaluation platform.

The contributions of this paper can be summarized in 3-fold:
\begin{itemize}

\item \textbf{Taxonomy of LLM Recommenders.}
We introduce a two-branch taxonomy that separates Pure and Augmented LLM Recommenders, offering a structured and comprehensive overview of the growing body of literature on refining LLM Recommenders.

\item \textbf{Unified Benchmark and Key Performance Factors.}
We establish a transparent, up-to-date benchmarking pipeline that standardizes data preprocessing and evaluation. By re-running representative models under identical conditions, we identify the specific design choices—such as identifier assignment, grounding strategies, and the use of collaborative signals, most strongly affect recommendation quality.

\item \textbf{Challenges and Research Roadmap.}
We summarize the key obstacles facing LLM Recommenders—such as distribution gap between language and recommendation semantics and the risk of propagating or amplifying bias. In addition, we articulate concrete directions for future work in this field.

\end{itemize}
For the following sections of this paper, we begin by a preliminary, in~\cref{sec:llm-recommender}, to specify our scope, followed by an overview of existing methods in~\cref{sec:overview_papers}.
Then, we show a dedicated experiment for representative approaches on a fair, unified platform to analyze the strengths of different methods in~\cref{sec:experiment}.
Finally, we discuss key challenges of LLM Recommenders in~\cref{sec:challenge} and highlight future research opportunities in~\cref{sec:future}.

\section{Preliminary}
\label{sec:llm-recommender}

LLM Recommenders utilize large language model for the recommendation tasks. In this paper, we focus on the models that leveraged LLM as the last decision-maker\footnote{That is, our scope does not include the use of LLMs only as auxiliary components in recommendation systems. Such cases represent a broader and distinct research territory, which we believe merits a separate survey and dedicated discussion.}. Formally, let $\mathcal{U}$ be the set of users, \iset be the set of items, \mset be the set of meta information and \rset be the set of recommendations.
The LLM Recommender $\mathbb{L}$ will map the given users, items and meta informations into a set of recommendations in \rset.
\begin{equation}
\label{eq:def_of_LLM_Recommender}
    \mathbb{L}: \uset \times \iset \times \mset \times f(\uset, \iset, \mset) \rightarrow \rset.
\end{equation}
where $f$ denotes the augmentation map, which can be any non-LLM techniques designed to improve the performance of the LLM Recommender  $\mathbb{L}$.

By examining the augmentation map $f$,
 we can further classify LLM Recommenders into two categories. The first is \textbf{Pure LLM Recommenders} (\cref{sec:llm-recommender:current_method:standalone}), which rely solely on the capabilities of LLMs without any external augmentation. That is, the augmentation map $f$ is a zero map. The second is \textbf{Augmented LLM Recommenders} (\cref{sec:llm-recommender:current_method:augmented}), which enhance LLMs with non-language model techniques such as conventional sequence models~\cite{Traditional:SASRec,Traditional:GRU4Rec}, where $f$ becomes a non-trivial map. In the following sections, we will examine both categories in greater detail.

\begin{table*}[t]
\centering
\small
{
\begin{tabular}{cp{4.3cm}p{8.5cm}}
\toprule
\bf Categories &  \bf Techniques & \bf Methods  \\
\midrule
\multirow{ 12 }{*}{\shortstack{Pure LLM \\Recommenders}} &
 Naive Embedding Utilization & \citet{Paper:17:BERT4Rec,Paper:66:LLM-Rec,Paper:64:RECFORMER,Paper:39}  \\
\cmidrule(lr){2-3}
& Naive Pretrained LM Finetuning
& \citet{Paper:2:P5,Paper:3:POD,Paper:29, Paper:4:RDRec} \\
\cmidrule(lr){2-3}
& Instruction Tuning
& \citet{Paper:7:TALLRec, Paper:10:GenRec, Paper:60, Paper:8} \\
\cmidrule(lr){2-3}
& Model Architectural Adaptations
& \citet{Paper:34, Paper:45, Paper:52:MoLoRec} \\
\cmidrule(lr){2-3}
&
Reflect-and-Rethink
& \citet{Paper:48, Paper:20:Re2llm, Paper:56:MACRec, Paper:59:RecPrompt, Paper:40, Paper:22:InstructRec, Paper:11,Paper:57,Paper:24:Instructagent} \\
\cmidrule(lr){2-3}
& Others &
\citet{Paper:1:CALRec, Paper:19, Paper:9:CLLM4Rec, Paper:32:LLM-TRSR, Paper:49, Paper:41, Paper:53} \\
\midrule
\multirow{ 6 }{*}{\shortstack{Augmented\\ LLM \\ Recommenders}} &
Semantic Identifiers Augmentation &
  \citet{Paper:46:P5_other_indexing, Paper:25:TIGER, Paper:12:LC-Rec, Paper:26:LIGER, Paper:47}   \\
\cmidrule(lr){2-3}
& Collaborative Modality Augmentation &
  \citet{Paper:62:CoLLM, Paper:31:Llara, Paper:30:iLoRA, Paper:63, Paper:42}   \\
\cmidrule(lr){2-3} & Prompts Augmentation
 & \citet{Paper:21:RPP,Paper:15:CRAG}     \\
\cmidrule(lr){2-3} & Retrieve-and-Rerank
 & \citet{Paper:6:NIR,Paper:16:Llamarec,Paper:18:Palr}     \\
\bottomrule
\end{tabular}
}
\caption{\textbf{Taxonomy of existing LLM Recommenders.} We categorize the methods into two general types: (1) Pure LLM Recommenders (\cref{sec:llm-recommender:current_method:standalone}), and (2) Augmented LLM Recommenders (\cref{sec:llm-recommender:current_method:augmented}). Within each category, methods are further grouped based on their underlying techniques and usage patterns.}

\label{tab:classification}
\end{table*}

\section{LLM Recommenders}
\label{sec:overview_papers}

\subsection{Pure LLM Recommenders}
\label{sec:llm-recommender:current_method:standalone}

\paragraph{Naive Embedding Utilization}
The most straightforward way to leverage LLMs in recommendation is to directly use their aggregated final state embeddings.
BERT4Rec~\cite{Paper:17:BERT4Rec} uses BERT~\cite{LLM:BERT} to represent each item as an embedding and computes the ranking score with a linear layer over the final hidden state.
Similarly, \citet{Paper:66:LLM-Rec} utilize final state embeddings to perform recommendations across multiple domains, enhanced by contrastive learning strategy.
RecFormer~\cite{Paper:64:RECFORMER} formulates an item as a 'sentence' and trains a transformer to comprehend the sentence sequence, subsequently retrieving the next sentence for item pool similarity search.
\citet{Paper:39} propose several methods to leverage LLM-derived embeddings, including using them for similarity-based retrieval, enhancing BERT4Rec through embedding initialization, and fine-tuning with session prompts.

\paragraph{Naive Pretrained LM Finetuning}
\label{sec:LLM:PLM_finetune}
While the aforementioned methods focus on exploiting LLM embeddings,
simultaneously, methods formulating recommendation as tasks and directly finetuning on pretrained LM have emerged. As a pioneer,
P5~\cite{Paper:2:P5} proposes whole-word embeddings, and fine-tunes pretrained T5 models~\cite{LLM:t5} to jointly perform multiple recommendation tasks within a single framework.
POD~\cite{Paper:3:POD} enhances the structure of P5 by distilling discrete prompts into continuous prompt embeddings.
Alternatively, \citet{Paper:29} enhance P5 by adding three auxiliary tasks and designing prompts that simplify input and output to improve item recognition.
Similarly, RDRec~\cite{Paper:4:RDRec} improves POD by distilling the rationale behind user-item interactions with Llama-2-7b~\cite{LLM:llama2} and adding them as auxiliary tasks.

\paragraph{Instruction Tuning}
As LLMs continue to scale up and instruction tuning starts to gain popularity,
TALLRec~\cite{Paper:7:TALLRec}, as a trailblazer, formulates user preference prediction as an instruction tuning task, where the model is guided by user historical preferred or unpreferred items.
GenRec~\cite{Paper:10:GenRec}, on the other hand, incorporates historical item titles into instructions to focus on next-item prediction.
BIGRec~\cite{Paper:60} further grounds the predicted items using the similarity of item embeddings, rather than relying solely on exact matching.
\citet{Paper:8} investigates rating prediction in zero-shot and few-shot settings, as well as through task tuning for classification or regression.

\paragraph{Model Architectural Adaptations}
In addition to standard applications of LLMs, numerous studies have proposed novel architectural adaptations of LLM backbones, specifically designed for recommendation systems.
LITE-LLM4REC~\cite{Paper:34} enhances the efficiency by introducing a hierarchical architecture that replaces token-based inputs with vectorized item embeddings and eliminates beam search with a direct scoring projection head.
RecPPT~\cite{Paper:45} reprograms a pretrained GPT-2
by aligning cross-domain item representations via a cross-attention-based prototype module, and injecting global user–item patterns through SVD-based initialization.
MoLoRec~\cite{Paper:52:MoLoRec} combines general and domain-specific recommendations. A genral domain LoRA~\cite{Others:LoRA} module is integrated with a specific one through parameters mixing.

\paragraph{Reflect-and-Rethink}
\label{sec:LLM:reasoning}
Beyond supervised learning, some methods improve LLM Recommenders by reflecting on outputs, refining prompts, or interpreting user intent to guide prompt design.
BiLLP~\cite{Paper:48} decomposes long-term recommendation into macro-learning for deriving high-level planning principles and micro-learning for grounding these plans into personalized actions.
Re2LLM~\cite{Paper:20:Re2llm} applies PPO training~\cite{Others:PPO} on textual information encoded by BERT, which then retrieves the appropriate hints generated by LLM for future ranking.
MACRec~\cite{Paper:56:MACRec} proposes a multi-agent collaborative framework in which agents with different roles work together to tackle specific recommendation tasks. Among them, a reflector agent is introduced to evaluate the correctness of the answers.
RecPrompt~\cite{Paper:59:RecPrompt} designs a pipeline to automatically perform prompt engineering, enhancing both the news recommender and the prompt optimizer.
LLM4ISR~\cite{Paper:40} prompts LLMs to semantically infer varying user intents in a session for next-item recommendation, and iteratively optimizes these prompts through self-reflection and UCB-based prompt selection.

Besides these reflective and prompt-centric strategies, several works focus more on capturing user preferences and intents through additional prompting or conversation with LLMs to enable more personalized recommendations.
InstructRec~\cite{Paper:22:InstructRec} uses GPT to annotate user preferences and intentions. It separates preferences, intentions, and tasks to design specific instructions for each recommendation task.
\citet{Paper:11} focus on conversational recommendation systems (CRS) and conduct a comprehensive study on zero-shot prompting.
LLM-ConvRec~\cite{Paper:57} prompts LLM to get the user intents, maintains a semi-structured dialogue state and performs retrieval-augmented recommendation and question answering via late fusion over item reviews.
iAgent~\cite{Paper:24:Instructagent}
incorporates user instructions and external knowledge. It parses the LLM to capture both internal and external user knowledge, dynamically updating the user profile via a LLM generator.

\paragraph{Others}
Other approaches focus on designing suitable training objectives, metadata summarization, data essence extraction, among others.
CALRec~\cite{Paper:1:CALRec} integrates the generation of next-item meta-information with hidden feature contrastive alignment as a combined training objective.
It leverages BM25~\cite{Others:bm25} as the retriever to extract similar items for grounding the generated outputs.
\citet{Paper:19} introduce Counterfactual Fine-Tuning (CFT), which adds a causal loss based on user behavior without prior interactions.
CLLM4Rec~\cite{Paper:9:CLLM4Rec}
encodes textual information into collaborative representations via jointly training a content LLM, which models user's review, and a collaborative LLM, which captures collaborative information.
LLM-TRSR~\cite{Paper:32:LLM-TRSR} handles text-rich histories
via hierarchical or recurrent summarization to capture the essence of user preferences.

LLM4Rerank~\cite{Paper:49} formulates the reranking process as a dynamically traversable function graph, where a zero-shot LLM sequentially selects and executes aspect-specific nodes.
TransRec~\cite{Paper:41} uses multi-facet identifiers (ID, title, attribute) and a position-free constrained generation mechanism via FM-index~\cite{Others:FM_index}, enabling accurate and efficient grounding.
DEALRec~\cite{Paper:53} focuses on the data side of the pipeline, as they aim to identify representative samples tailored for few-shot finetuning of LLMs.

\subsection{Augmented LLM Recommenders}
\label{sec:llm-recommender:current_method:augmented}

Recently, there has been growing interest in methods that augment LLM Recommenders by incorporating non-LLM techniques. By examining the augmentation map $f$, we can further categorize into the following classes.

\paragraph{Semantic Identifiers Augmentation}
Semantic Identifiers (or Semantic IDs) augmentation methods represent user or item IDs as implicit semantic sequences with the help of auxiliary coding techniques. For example, \citet{Paper:46:P5_other_indexing} explore various identifier strategies for the P5 structure, such as Collaborative Indexing (CID), which leverages Spectral Matrix Factorization~\cite{Others:spectral_mf} to cluster items and expresses Item IDs as a sequence of extra tokens based on clustering results.

Since late 2023, RQ-VAE-based~\cite{Others:RQ-VAE} approaches have gained significant attention. A prominent example is TIGER~\cite{Paper:25:TIGER}, which uses Sentence-T5~\cite{Others:sentence-t5} to encode item metadata into latent embeddings, followed by RQ-VAE to produce hierarchical discrete codes. Each RQ code is treated as a token, with a hashed user ID token prepended to the input sequence. Another example is LC-Rec~\cite{Paper:12:LC-Rec}, which jointly trains the introduced RQ-VAE-based tokens with languages tokens via instruction tuning. LIGER~\cite{Paper:26:LIGER} proposes a hybrid method combining dense and generative models, with a focus on cold-start queries. LETTER~\cite{Paper:47} further incorporates additional signals to enhance the pure textual semantics IDs.

\paragraph{Collaborative Modality Augmentation}
Collaborative Modality Augmentation methods seek to align collaborative information with language, usually by projecting embeddings derived from traditional collaborative models into the language space.
CoLLM~\cite{Paper:62:CoLLM} and LLaRA~\cite{Paper:31:Llara} are two representative examples that integrate collaborative embeddings
into language prompts. CoLLM constructs task-specific prompts to inject collaborative information, while LLaRA concatenates projected collaborative embeddings with item titles.
\citet{Paper:30:iLoRA} propose iLoRA, which improves LLaRA by mitigating gradient misalignment through a linear combination of multiple expert LoRA modules, weighted based on outputs from conventional collaborative models.
A-LLMRec~\cite{Paper:63} aligns collaborative and item textual information with SASRec and SBERT~\cite{Others:SBERT}.
The aligned embeddings are then
projected onto the frozen LLM to perform recommendations.
Llama4Rec~\cite{Paper:42} integrates information from traditional recommenders into the instruction-tuning data for LLMs, and fuses outputs from both models using a long-tail-aware strategy to enhance performance.

\paragraph{Prompts Augmentation}
Prompts augmentation methods utilize non-LLM techniques to improve the quality of prompts.
RPP~\cite{Paper:21:RPP} dynamically adapts and updates prompts for reranking task through an actor-critic framework, operating over four defined prompting segments. The actor and critic models are implemented with the help of BERT and GRU~\cite{Others:GRU} to facilitate policy updates.
CRAG~\cite{Paper:15:CRAG} improves \citet{Paper:11} for CRS scenario by context-aware collaborative retreival and reflect-and-rerank mechanism.

\paragraph{Retrieve-and-Rerank}
\label{sec:llm-recommender:current_method:augmented:cand}
A number of methods retrieve top-ranked candidates from non-LLM techniques and rerank via LLM Recommenders.
NIR~\cite{Paper:6:NIR} employs multi-hot vector representations to retrieve similar users or items for the construction of candidate sets. Then, a three-step prompting is used to rerank.
LlamaRec~\cite{Paper:16:Llamarec} first retrieves candidate items using LRURec~\cite{non_Paper:LRURec}, and subsequently reranks them through instruction tuning and a simple verbalization strategy that maps vocabulary to labels.
PALR~\cite{Paper:18:Palr} also retrieves candidate items using any robust recommendation method, followed by reranking. Additionally, it leverages LLMs to summarize user preferences and incorporates them into prompts for the finetuning stage.

\begin{table*}[t]
  \centering
  \small
  \begin{tabular}{lcccc|cccc}
    \toprule
    \multirow{2}{*}{\textbf{Method}} &
    \multicolumn{4}{c|}{\textbf{Musical Instruments}} &
    \multicolumn{4}{c}{\textbf{Industrial and Scientific}} \\
    & \textbf{Hit@5} & \textbf{NDCG@5} & \textbf{Hit@10} & \textbf{NDCG@10}
      & \textbf{Hit@5} & \textbf{NDCG@5} & \textbf{Hit@10} & \textbf{NDCG@10} \\
    \midrule
    \multicolumn{9}{l}{\it Traditional}\\
    SASRec
       & .0224 & .0117 & .0379 & .0167 & .0152 & .0087 & .0255 & .0120 \\
    GRU4Rec
       & .0203 & .0133 & .0322 & .0171 & .0171 & .0118 & .0256 & .0145 \\
    \midrule
    \multicolumn{9}{l}{\it Pure LLM}\\
    P5
      & .0159 & .0108 & .0239 & .0134 & .0139 & .0093 & .0211 & .0116 \\
    POD
      & .0142 & .0096 & .0213 & .0119 & .0133 & .0082 & .0195 & .0102 \\
    RDRec
      & .0172 & .0114 & .0260 & .0142 & .0139 & .0086 & .0205 & .0107 \\
    GenRec
      & .0152 & .0094 & .0227 & .0119 & .0084 & .0046 & .0119 & .0058 \\
    BIGRec
      & .0236 & .0133 & .0420 & .0192 & .0160 & .0107 & .0280 & .0144 \\
    \midrule
    \multicolumn{9}{l}{\it Augmented LLM}\\
    P5-CID
      & .0283 & .0180 & .0451 & .0234 & .0137 & .0089 & .0205 & .0110 \\
    TIGER
      & .0332 & .0216 & .0517 & .0276 & .0241 & .0158 & .0385 & .0204 \\
    LETTER-TIGER
      & .0339 & .0224 & .0521 & .0282 & .0256 & .0165 & .0396 & .0210 \\
    \bottomrule
  \end{tabular}
  \caption{\textbf{Results of sequential recommendation.}}
  \label{tab:seq_exp_results}
\end{table*}

\section{Experiments}
\label{sec:experiment}
Although many benchmarks exist for recommender systems, there remains a lack of comprehensive comparisons between \emph{Pure} and \emph{Augmented} LLM Recommenders under consistent, fair, and modern evaluation settings. To fill this gap, we design a unified experimental framework and use it to systematically assess the performance of both categories.
This section begins by describing the dataset preprocessing and evaluation tasks, followed by experiments that highlight the strengths and limitations of each approach under standardized conditions.

\subsection{Experimental Setup}
Many recommender system studies employ custom preprocessing steps, making it challenging to compare results across different methods.\footnote{For example, models like P5~\cite{Paper:2:P5} and POD~\cite{Paper:3:POD} commonly preprocess user histories by sequentially assigning numerical IDs to items. The first user's earliest interaction might receive ID '1', the next unseen item ID '2', and so forth. Due to the extensive pretrained knowledge embedded in LLMs, this approach can inadvertently introduce information leakage, as noted by TIGER~\cite{Paper:25:TIGER}.} Additionally, existing evaluations often utilize outdated datasets such as Amazon'18~\cite{Dataset:Amazon_18} or MovieLens100K~\cite{Dataset:Movielens}, which may no longer represent contemporary user-item interaction patterns accurately.

\subsubsection{Dataset Preparation}
We adopt the Amazon'23 dataset~\cite{Dataset:Amazon_23}, capturing up-to-date user-item interactions. Among available categories, we select \textit{Musical Instruments} and \textit{Industrial and Scientific} because their user-item scales align with mainstream recommenders. Following standard procedures, we implement a 5-core setting\footnote{The 5-core setting filters out users and items with fewer than five interactions, ensuring sufficient data density.}, and apply a leave-one-out evaluation protocol\footnote{We use the second-to-last interaction for validation and the last interaction for testing.}. To maintain consistency and fairness across experiments, methods that use naive numerical identifiers---specifically P5~\cite{Paper:2:P5}, POD~\cite{Paper:3:POD}, RDRec~\cite{Paper:4:RDRec}---receive randomly assigned user and item IDs within the ranges \([0, \#\text{Users} - 1]\) and \([0, \#\text{Items} - 1]\), respectively. Detailed dataset statistics and preprocessing are provided in~\cref{sec:appendix_data_stats}.

\subsubsection{Sequential Recommendation}
\label{sec:exp:seq}
We now introduce the primary recommender system task\footnote{We define more recommendation tasks in \cref{sec:appendix_more_recsys_tasks}.} examined: Sequential Recommendation. Given a user $u \in \mathcal{U}$ and their time-ordered interaction history $h_u = (i_{h_1}, i_{h_2}, i_{h_3}, \ldots, i_{h_{n-1}})$, where each $i_{h_j} \in \mathcal{I}$, along with optional user and item metadata $m_u \in \mathcal{M}$ and $m_{h_u} \subset \mathcal{M}$, the goal of the Sequential Recommendation model, denoted as $\mathbb{L}_{S}$, is to predict the next item $i_{h_n} \in \mathcal{I}$ that the user is likely to interact with. To be more specific, the problem can be defined as:
{
\begin{equation}
\label{eq:SRS_model}
\footnotesize
\begin{aligned}
    \mathbb{L}_{S}(u, h_u, m_u, m_{h_u}, f_u^{h_u}, k)
    = \{i_{p_j} \, | \,  i_{p_j} \in \mathcal{I},  \, j=1 \ldots k\}.
\end{aligned}
\end{equation}
}

Here, $f_u^{h_u} = f(u, h_u, m_u, m_{h_u})$ denotes the augmented information after non-LLM enhancement map applied and $k$ denotes the number of returned items. The ground truth next item $i_{h_n}$ is expected to be one of the predicted items $i_{p}$.
\begin{figure*}[t]
  \centering
  \includegraphics[width=1.0\textwidth]{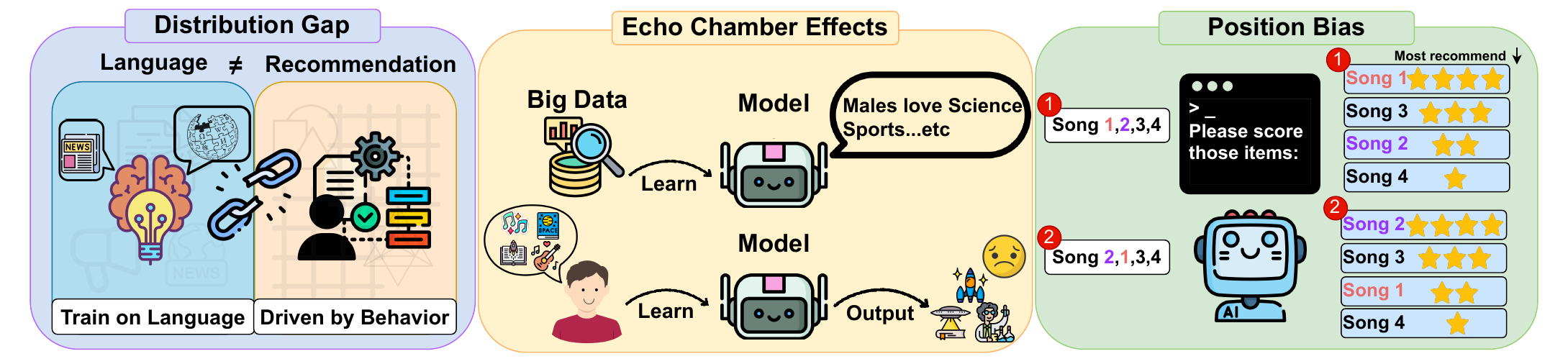}
  \vspace{-2mm}
  \caption{\textbf{Illustration of the challenges in LLM Recommenders}--- Distribution Gap between Recommendation and Language Semantics, Echo Chamber Effects, and Position Bias.}
  \label{fig:challenges}
\end{figure*}

\subsection{Performance Analysis on Sequential Recommendation}
We evaluate the performance of Pure and Augmented LLM Recommenders and present the results of NDCG@K and HIT@K in~\cref{tab:seq_exp_results}. For clearer comparison, we include two traditional (non-LLM) recommender methods as baselines:
\begin{compactitem}
\item \textbf{SASRec~\cite{Traditional:SASRec}:} A transformer-based sequential recommendation model that captures user preferences through self-attention mechanisms.
\item \textbf{GRU4Rec~\cite{Traditional:GRU4Rec}:} An RNN-based model utilizing gated recurrent units (GRUs) to model user sequential behavior.
\end{compactitem}
The model implementation details are documented in~\cref{sec:appendix_model_implementation_details}.

\paragraph{Pure LLM Recommenders}
For Pure LLM Recommenders, item representation plays a critical role in overall performance. GenRec~\cite{Paper:10:GenRec}, for instance, uses item titles as item representations and directly generates the next item in natural language form. In contrast, models such as P5~\cite{Paper:2:P5}, POD~\cite{Paper:3:POD}, and RDRec~\cite{Paper:4:RDRec} adopt numerical item representations for training and prediction. This reduces the risk of generating incorrect or non-existent item titles, which is more likely when relying solely on textual representations. Notably, while BIGRec~\cite{Paper:60} also uses item titles for item representation, it addresses the risk of generating invalid items by mapping the generated output into an embedding space and performing similarity search within the item pool. This grounding strategy helps align outputs with real items, thereby improving recommendation performance.

\paragraph{Augmented LLM Recommenders}
In contrast to naive numerical item indexing, TIGER~\cite{Paper:25:TIGER} enhances item representations by leveraging metadata to pre-cluster items and create Semantic IDs. These IDs encode richer information than plain numeric indices, leading to significant performance improvements. Further gains can be achieved by incorporating collaborative signals. P5, for example, shows improved performance when using CID~\cite{Paper:46:P5_other_indexing} representations that embed collaborative information. LETTER-TIGER~\cite{Paper:47} similarly extends TIGER by incorporating collaborative signals during the item clustering process, enabling Semantic IDs to reflect collaborative patterns and further enhancing model effectiveness. Overall, we observe that Augmented LLM Recommenders generally outperform traditional recommender baselines and even Pure LLM Recommenders in our experiment scenarios.

\section{Challenges of LLM Recommenders}
In this section, we discuss key challenges faced by LLM Recommenders.~\cref{fig:challenges} provides a visual summary of these challenges.
\label{sec:challenge}

\paragraph{Distribution Gap between Recommendation and Language Semantics}
\label{sec:challenge:distribution_gap}
The goal of recommendation systems is to provide accurate suggestions based on collaborative information, such as user-item interaction patterns.  To achieve this, it is essential for recommenders to effectively model user underlying behavior. LLMs, trained on vast text corpora, are expected to implicitly encode some aspects of such patterns.
However, recent research has shown that directly leveraging the implicit collaborative knowledge within LLMs remains a challenge.
~\citet{Paper:5:LLMRank}, ~\citet{Paper:7:TALLRec}, ~\citet{Paper:8} highlight this issue by showing that, under standard zero-shot prompting, LLMs often fail to capture the temporal order of user interactions.
Moreover, traditional methods such as Matrix Factorization~\cite{Others:BPR} or Multi-Layer Perceptrons~\cite{Others:MLP} can sometimes outperform large LLMs when relying solely on zero-shot or few-shot prompting.
This issue is often addressed by tuning LLMs to better comprehend the underlying objectives of recommendation tasks (\cref{sec:LLM:PLM_finetune}).

Even with exhaustive tuning, LLM Recommenders may still be influenced by the pretrained language semantics\footnote{As highlighted in CoLLM~\cite{Paper:62:CoLLM}, for example, two items with similar item descriptions may still be consumed by different kinds of users.}. This can prevent LLMs from faithfully capturing the true collaborative semantics.
To address this issue, recent methods typically construct alignment tasks to help LLMs understand and incorporate collaborative or statistical information. Specifically, some approaches aim to align language tokens with collaborative tokens by jointly training both token types~\cite{Paper:9:CLLM4Rec, Paper:12:LC-Rec, Paper:42}, while others train a projection module that maps collaborative tokens into the language token space~\cite{Paper:31:Llara, Paper:62:CoLLM, Paper:63}.

\paragraph{Echo Chamber Effects}
\label{sec:challenge:echo_chamber}
The echo chamber effect refers to a situation in which individuals are predominantly exposed to information that reinforces models' preexisting beliefs, often due to selective exposure, algorithmic filtering, or even the underlying social biases inherent in LLMs. In recommender systems, this can result in users repeatedly receiving a narrow range of items, irrespective of their current intent.
CFaiRLLM~\cite{Paper:13:CFaiRLLM} argues that LLMs inherently exhibit social biases, originating from unregulated training data. To assess this, they propose an enhanced evaluation framework for fairness, focusing on social biases such as gender, race, and age.
Conversational Recommender Systems (CRS)~\cite{Paper:11, Paper:24:Instructagent, Paper:57} aim to mitigate this issue by enabling users to interact with LLMs through dialogue, allowing the model to infer the user's true intent and provide more accurate recommendations. Additionally, reinforcement learning methods such as Re2LLM~\cite{Paper:20:Re2llm} are designed to reflect and explore the user's underlying preferences.

\paragraph{Position Bias}
\label{sec:challenge:position_bias}
Position bias refers to the tendency for the perceived relevance or importance of recommended items to be influenced by their position in the prompt input list, which should ideally yield symmetric outputs under permutations. In recommendation systems, especially in zero-shot prompting scenarios, the position of the ground-truth item within the candidate set is significantly affected by this bias~\cite{Paper:43,Paper:5:LLMRank}. To address this issue, existing approaches either employ bootstrapping techniques, such as repeatedly shuffling and reranking the candidate set~\cite{Paper:5:LLMRank}, or construct a transition matrix based on item positions to postprocess the prior probability distribution~\cite{Paper:43}. Notably, position bias also arises in other LLM tasks ~\cite{hofstatter2021mitigating,Others:position_bias:LLM_judge,wei-etal-2024-unveiling}, just to name a few, and should be carefully considered when designing LLM-based recommendation systems, particularly for reranking tasks.

\section{Future Directions}
\label{sec:future}
Cold-Start and Cross-Domain Generalizability are long-standing challenges in recommendation~\cite{10.1145/3460231.3474228, 10.1145/3637528.3671799}.
LLM Recommenders offer a promising solution due to their ability to understand rich textual metadata. By semantically analyzing item descriptions, tags, and reviews, these models can group similar items, even unseen ones, enabling implicit clustering, without relying heavily on user interaction history.

\paragraph{Cold-Start}
Cold-Start refers to users or items with extremely sparse interactions in user-item patterns. Conventional collaborative methods mainly rely primarily on behavioral data, making it difficult to estimate similarity or preference signals in such cases. To mitigate the cold-start issue, mainstream LLM Recommenders~\cite{Paper:25:TIGER,Paper:12:LC-Rec,Paper:47} commonly leverage metadata to pre-cluster semantically similar items. As demonstrated in TIGER~\cite{Paper:25:TIGER}, the use of LLMs enables recommendation in cold-start scenarios, offering promising potential. However, as discussed in LIGER~\cite{Paper:26:LIGER}, current generative recommenders still face challenges in these settings. The conditional probabilities objective of decoder tends to overfit to items seen during training, leading to a significantly reduced capability to generate cold-start items.
A related concern is whether incorporating collaborative signals may degrade performance as collaborative filtering based methods tends to suffer more from cold-start scenarios. In our survey, we found a lack of studies that systematically evaluated this issue.
Hence, further research is still needed to systematically evaluate and address the aforementioned problems across both types of LLM Recommenders.

\paragraph{Cross-Domain Generalizability}
Cross-Domain Generalizability refers to a model's ability to perform effectively across different domains. It becomes challenging when a model trained in one domain encounters different attributes and user-intent patterns in another domain, resulting in distributional misalignment. The ability of LLMs to recognize high-level concepts enables recommendation systems to transfer behavioral patterns across domains, opening up new possibilities for cross-domain recommendation. Recent efforts~\cite{Paper:66:LLM-Rec,Paper:52:MoLoRec,Paper:45} have explored
through few approaches such as direct fine-tuning, plug-and-play modules, and model reprogramming. This direction remains largely underexplored, with relatively few studies addressing and analyzing this issue, leaving ample room for future research.

\section{Conclusion}
\label{sec:con}

In alignment with the current trend of integrating LLMs into recommendation system, we propose a two-branch taxonomy for LLM Recommenders, by classifying them into Pure and Augmented methods. Under this framework, we provide a comprehensive overview of existing models and systematically evaluate representative ones on a fair, standardized, and up-to-date benchmarking pipeline. Finally, we identify key challenges, misalignment between recommendation and language semantics, the echo chamber effect, and positional bias, as well as highlight promising capabilities, cold-start handling and cross-domain generalizability, to inform future directions. We hope this comprehensive review will serve as a catalyst for advancing research in the LLM Recommenders community.

\section*{Limitations}
The definition of LLM-based recommenders varies across the literature, depending on the role and integration level of LLMs within the recommendation pipeline. In this work, we explicitly focus on scenarios where LLMs function as the final decision maker, that is, they directly generate or select recommendations. As such, methods that employ LLMs in auxiliary roles, such as data preprocessing, feature extraction, prompt generation, or candidate reranking, are excluded from our taxonomy and discussion.
This narrowed scope allows for a clearer comparison of LLMs when positioned as core recommendation engines, but it also means that our analysis does not cover hybrid or support-role use cases, which may involve important yet indirect contributions from LLMs.

Moreover, our discussion centers on the unique challenges and design considerations that emerge specifically from using LLMs as recommenders. We do not aim to address broader issues common to all recommender systems, unless they manifest in a distinct way due to the use of LLMs.

\section*{Acknowledgments}

We thank for NTU Web Mining and Information Retrieval Lab for the provided computational resources. Furthermore, we also thank for Tzu-Lin Kuo and Tzu-Wei Chiu from National Taiwan University for their discussion and wonderful advice, and acknowledge the use of ChatGPT, developed by OpenAI, which was used to polish the language of the manuscript.

\bibliography{custom}

\appendix
\section{Data Statistics and Preparation Details}
\label{sec:appendix_data_stats}

\begin{table*}[h]
    \centering
    \caption{Dataset statistics for the two chosen Amazon'23 benchmarks. The \textbf{History Length} is the historical interaction per user.}
    \begin{tabular}{lcccc}
        \toprule
        \textbf{Dataset} & \textbf{\# Users} & \textbf{\# Items} & \multicolumn{2}{c}{\textbf{History Length}} \\
        \cmidrule(lr){4-5}
                          &                   &                   & \textbf{Mean} & \textbf{Median} \\
        \midrule
        Musical Instruments  & 57,439 & 24,587 & 8.91  & 7 \\
         Industrial \& Scientifc  & 50,985  & 25,848 & 8.10 & 6 \\
        \bottomrule
    \end{tabular}
    \label{tab:dataset_stats}
\end{table*}

In this session, we showcase more on our dataset preparation, as well as the dataset statistics.

We preprocessed the \textit{Musical\_Instruments} and \textit{Industrial\_and\_Scientific} datasets from the Amazon'23 dataset website\footnote{\url{https://amazon-reviews-2023.github.io/index.html}}~\cite{Dataset:Amazon_23}. We downloaded both the review data and the metadata, along with the preprocessed 5-Core leave-one-out split. Next, we removed any item metadata entries that were not present in all splits of the 5-Core dataset.

As described in~\cref{sec:experiment}, we randomly assigned a unique naive numerical identifier to each user and item. This step is particularly for methods, P5~\cite{Paper:2:P5}, POD~\cite{Paper:3:POD} and RDRec~\cite{Paper:4:RDRec}, which rely on such naive numerical ID representations. We also merged the 5-Core data with the review data by matching on the same \textit{user\_id}, \textit{parent\_asin}, and \textit{timestamp}.
All above-mentioned preprocessing procedures were finalized prior to conducting any experiments.
The statistics of the final preprocessed datasets are presented in~\cref{tab:dataset_stats}.

\section{Model Implementation Details}
\label{sec:appendix_model_implementation_details}

In this section, we provide details on the model implementations, the source code used, and the modifications made to adapt to our setup. We aim to share these details to facilitate reproducibility.

\begin{itemize}

    \item \textbf{SASRec}~\cite{Traditional:SASRec}: We used the \textit{PyTorch} implementation provided by the author of \textit{S3Rec}\footnote{\url{https://github.com/RUCAIBox/CIKM2020-S3Rec}}~\cite{Traditional:S3Rec}. The original implementation filters out items that the user has previously interacted with during prediction, which is not a guarantee of common LLM Recommenders. To ensure a fair comparison across models, we removed this filtering setting.

    \item \textbf{GRU4Rec}~\cite{Traditional:GRU4Rec}: We used the official Pytorch Implementation\footnote{\url{https://github.com/hidasib/GRU4Rec_PyTorch_Official}}. To adapt the original GRU4Rec method to our dataset, we modified both the data preprocessing and evaluation pipeline to match our leave-one-out protocol.
    Moreover, we customized the original evaluation logic to focus only on predicting the final item of each session, replacing GRU4Rec's default step-by-step evaluation with the leave-one-out evaluation strategy.

    \item \textbf{P5}~\cite{Paper:2:P5}: We used the official implementation of P5\footnote{\url{https://github.com/jeykigung/P5}} and further preprocessed the text in each review to match the explanation format expected by P5. Specifically, we extracted the text from the beginning of the review up to the first occurrence of a period ('.'), treating it as a single sentence. This preprocessing strategy closely follows the example provided by the authors\footnote{\url{https://github.com/jeykigung/P5/blob/main/preprocess/data_preprocess_amazon.ipynb}, shown in the output of cell [13].}. For naive numerical ID, to ensure a fair comparison, we used our own randomly assigned IDs as described in Section~\ref{sec:experiment}.

    \item \textbf{POD}~\cite{Paper:3:POD}: We used the official code implementation\footnote{\url{https://github.com/lileipisces/POD}} and adapted portions of the codes, mainly generation-related, to reconcile the discrepancies between our computational environments. After these modifications, we verified that we could reproduce the results on the \textit{Beauty} benchmark reported in the original paper. Other preprocessing strategies, such as explanation extraction and naive numerical ID assignment, were applied in the same manner as in our implementation of P5.

   \item \textbf{RDRec}~\cite{Paper:4:RDRec}:We used the official codes\footnote{\url{https://github.com/WangXFng/RDRec}} for training and testing. The explanation preprocessing and naive numerical ID assignment were applied in the same manner as in our implementations of P5 and POD. For the distillation step, to improve throughput, we employed \texttt{vLLM}~\cite{Others:vllm} on \texttt{LLaMA2-7B}~\cite{LLM:llama2}, which is the same LLM in the paper, to obtain the rationales.

    \item \textbf{GenRec}~\cite{Paper:10:GenRec}: We used the official codes\footnote{\url{https://github.com/rutgerswiselab/GenRec}} for training and randomly sampled 50,000 reviews from the training set to serve as the training data. Due to the relatively slow throughput of the \texttt{generate} function in the \texttt{Transformers} library, we replaced the original inference code with \texttt{vLLM} to accelerate generation.

    \item \textbf{BIGRec}\cite{Paper:60}: We utilize the official implementation provided by the authors for training, evaluation, and testing.\footnote{\url{https://github.com/SAI990323/BIGRec}} To enhance inference throughput, we employ \texttt{vLLM} for acceleration. Additionally, the SASRec model, which is used for combining output scores, is trained by us.
    For more details, please refer to the previous explanation of SASRec implementation details.

    \item \textbf{P5-CID}~\cite{Paper:46:P5_other_indexing}: We adopted the official codebase\footnote{\url{https://github.com/Wenyueh/LLM-RecSys-ID}} and followed the data transformation procedure outlined in the repository to restructure the input format of our dataset. The remainder of the training pipeline was kept consistent with the original CID setup.

    \item \textbf{TIGER}~\cite{Paper:25:TIGER}: We use the codebase from LETTER\footnote{\url{https://github.com/HonghuiBao2000/LETTER}}~\cite{Paper:47}, as the original implementation of TIGER~\cite{Paper:25:TIGER} is not publicly available. To align with TIGER's settings, we remove enhancements introduced by LETTER to the RQ-VAE, including collaborative regularization, diversity regularization, and Sinkhorn-Knopp Algorithm. After training the RQ-VAE, we follow the remainder of LETTER’s pipeline, proceeding with transformer training.

    \item \textbf{LETTER-TIGER}~\cite{Paper:47}: LETTER-TIGER is one of the LETTER enhanced models. We follow the original pipeline\footnotemark[18] that requires generating item-tokenized sequences via a pretrained SASRec model and a learned discrete codebook (RQ-VAE). Since the authors did not release the pretrained collaborative filtering model, we reproduced it using SASRec on our dataset and extracted the item embeddings as \texttt{*.pt} files. We then quantized item representations through RQ-VAE. To cope with hardware constraints, we loaded the pretrained language model specified in the original paper, \texttt{LLaMA2-7B}, using half-precision (float16) during embedding generation. All other components were implemented following the official repository.

\end{itemize}

\section{More on Recommendation Tasks}
\label{sec:appendix_more_recsys_tasks}

This appendix section aims to introduce more common recommendation tasks besides from Sequential Recommendation (\cref{sec:exp:seq}).

\paragraph{Reranking Recommendation}
Given a user $u$, a set of $n$ candidate next items $c_u = \{i_{c_j} \mid i_{c_j} \in \mathcal{I}, \, j = 1...n\}$ and optional metadata, $m_u$ and $m_{c_u}$, including the user time-ordered interaction history, the goal of the reranking model is to reorder the candidates based on their relevance to the user. Formally, the reranker model, $\mathbb{L}_{re}$, aims to select the top-$k$ items from $C_u$ as follows:

{\small
\begin{equation}\label{eq:reranking_model}
\begin{aligned}
    \text{\normalsize $\mathbb{L}_{re}$}(u, c_u, m_u, m_{c_u}, f_u^{c_u}, k) = \{i_{p_j} | i_{p_j} \in c_u,\, j = 1...k, \\
    \text{Rank\_Score}(i_{p_j}) \geq \text{Rank\_Score}(i_{p_{j+1}})\}.
\end{aligned}
\end{equation}
}Here, $k \leq n$ denotes the number of top-ranked items to be returned.

\paragraph{Discussion on Reranking Task}

Mainstream LLM Recommenders typically construct candidate sets using either (1) \emph{random negative sampling}, where the ground-truth item $i_{h_n}$ is included among randomly sampled negatives~\cite{Paper:2:P5,Paper:3:POD,Paper:31:Llara}, or (2) \emph{retrieve-and-rerank} (\cref{sec:llm-recommender:current_method:augmented:cand}), where candidates are retrieved from traditional recommenders or filtered sets without guaranteeing the presence of the positive item~\cite{Paper:16:Llamarec,Paper:6:NIR,Paper:18:Palr}.
While random sampling provides a fair reranking benchmark, its negatives may lack relevance, limiting real-world applicability. In contrast,
 the retrieve-and-rerank paradigm offers a more realistic evaluation setup but suffers from the drawback of that the positive item is not guaranteed to be present in the candidate set, making it difficult to truly evaluate the ability of LLM Reranker. Current benchmark tends to combine retrieve and rerank stages and compare the performance to methods dedicated to sequential recommendation task (\cref{sec:exp:seq}).

While not unique to LLM Recommenders, these issues suggest opportunities for improving benchmark design in terms of fairness and comprehensiveness.
Future work may also focus on establishing a more holistic evaluation framework—one that balances realism with consistency and enables comprehensive comparisons across diverse recommendation scenarios.

\paragraph{Others}

In addition to the most common considered recommendation tasks, sequential and reranking recommendation,
several modern LLM Recommenders have been extended to support a variety of task scenarios.

These include \textbf{Binary Recommendation}, which aims to predict the user preferences toward a specific item. The outcome is either \textit{"Like"} or \textit{"Dislike"}.
Similarly, \textbf{Rating Prediction} is an extensive one by further predicting the rating of the proposed item.
Often, ground-truth binary preferences are derived from ratings using a predefined threshold $b_t$~\cite{Paper:7:TALLRec}: a ground truth rating $r_{\text{gt}}$ is treated as a positive instance ("Yes") if $r_{\text{gt}} \geq b_t$, and as a negative instance ("No") otherwise.

Other emerging tasks include \textbf{explanation generation}, where the system generates textual reviews, and \textbf{conversational recommendation}, which involves generating recommendations through interactive dialogue between the user and the recommenders.

\section{Survey Methodology}
To ensure a comprehensive analysis of the rapidly evolving field of LLM Recommenders, we adopt a structured survey methodology encompassing scope definition, literature collection, paper categorization, and empirical evaluation.

We begin by clearly defining the scope of our study. This work specifically focuses on LLM acting as the final decision maker (\cref{sec:llm-recommender}). We intentionally exclude approaches where LLMs are used solely for auxiliary functions, such as feature extraction, data extraction, and others that lie beyond the scope of our taxonomy.

To gather relevant literature, we performed systematic searches across Google Scholar, Semantic Scholar, the ACL Anthology, and major conference proceedings including ACL, EMNLP, NAACL, SIGIR, RecSys, WWW, and NeurIPS. Search terms included "LLM Recommender," "Prompt-based Recommendation," "Generative Recommendation," and related variants. We also employed backward and forward citation tracing, beginning from seminal works such as P5, TIGER, and TALLRec. This process yielded a broad set of papers, from which we excluded those outside the scope of our study, resulting in a curated collection for in-depth discussion and analysis.

Next, we examine each paper in detail and broadly categorize them into two designed groups, Pure LLM Recommenders and Augmented LLM Recommenders.
For Pure LLM Recommenders (\cref{sec:llm-recommender:current_method:standalone}), we focus on how the tasks and models are adapted to leverage LLMs. Based on these adaptations, we classified the papers into Naive Embedding Utilization, Naive Pretrained Language Model Finetuning, Instruction Tuning, Model Architectural Adaptations, Reflect-and-Rethink, and Others.
On the other hand, the classification of Augmented LLM Recommenders (\cref{sec:llm-recommender:current_method:augmented}) emphasizes how non-LLM techniques are integrated into LLM Recommeders. Accordingly, we categorize the relevant papers into Semantic Identifiers Augmentation, Collaborative Modality Augmentation, Prompts Augmentation, and Retrieve-and-Rerank.

To provide deeper insights between the two categories, we also perform an empirical comparison of representative models on Amazon'23 dataset~\cite{Dataset:Amazon_23}.  These results, presented in~\cref{sec:experiment}, offer empirical insights under our experiment settings.

\end{document}